\documentclass[aps,eqsecnum,preprint,floats,epsf,epsfig,nofootinbib]{revtex4}

\usepackage{graphicx}

\usepackage{bm}

\newcommand{\beq}{\begin{equation}}
\newcommand{\eeq}{\end{equation}}
\newcommand{\beqs}{\begin{eqnarray}}
\newcommand{\eeqs}{\end{eqnarray}}
\newcommand{\lsim}{\mathrel{\raisebox{-
.6ex}{$\stackrel{\textstyle<}{\sim}$}}}

\begin{document}

\font\el=cmbx10 scaled \magstep2{\obeylines\hfill October, 2011}

\vskip 1.5 cm

\centerline{\large\bf Revisiting Axial-Vector Meson Mixing}
\bigskip
\bigskip
\centerline{\bf Hai-Yang Cheng}
\medskip
\centerline{Institute of Physics, Academia Sinica}
\centerline{Taipei, Taiwan 115, Republic of China}
\medskip
\centerline{and}
\medskip
\centerline{C.N. Yang Institute for Theoretical Physics, Stony Brook University}
\centerline{Stony Brook, NY 11794}

\bigskip
\bigskip
\centerline{\bf Abstract}
\bigskip
\small

Various phenomenological studies indicate that the mixing angle $\theta_{K_1}$ of $K_{1A}$ and $K_{1B}$, the strange partners of the axial-vector mesons $a_1(1260)$ and $b_1(1235)$, respectively, lies in the vicinity of $35^\circ$ or $55^\circ$, but whether this angle is larger or smaller than $45^\circ$ still remains controversial.
When the $f_1(1285)$-$f_1(1420)$ mixing angle $\theta_{^3P_1}$ and the $h_1(1170)$-$h_1(1380)$ mixing angle $\theta_{^1P_1}$ are determined from the mass relations, they depend on the masses of $K_{1A}$ and $K_{1B}$, which in turn depend on the mixing angle $\theta_{K_1}$.
We show that the approximate decoupling of the light $q\bar q$ state from the heavier $s \bar s$ state, which is empirically valid for vector, tensor and $3^{--}$ mesons, when applied to isoscalar axial-vector mesons, will enable us to discriminate different solutions of $\theta_{^3P_1}$ and $\theta_{^1P_1}$ and pick up $\theta_{K_1}\sim 35^\circ$. Indeed, for $\theta_{K_1}\sim 55^\circ$, the predicted $\theta_{^1P_1}$ disagrees sharply with the recent lattice calculation and the implied large $s\bar s$ content of $h_1(1170)$ and $q\bar q$ component of $h_1(1380)$ cannot explain the observation of their strong decays.
We conclude that $\theta_{K_1}$ is smaller than $45^\circ$.

\pagebreak

\section{Introduction}

The mixing of self-conjugate mesons in a generalized QCD-like theory was recently discussed in \cite{CS} with emphasis on the role of decoupling.
The mixing of the flavor-SU(3) singlet and octet states of vector and tensor
mesons to form mass eigenstates is of fundamental importance in hadronic
physics. In the case of the vector mesons, the physical $\omega$ is mostly
comprised of the isospin-singlet combination $(u \bar u + d \bar d)/\sqrt{2}$,
while $\phi$ is mostly an $s \bar s$ state. In a modern context, some insight into this comes from the Appelquist-Carazzone
decoupling theorem \cite{ac}, according to which, in a vectorial theory, as the mass of a particle gets large compared with a relevant scale, say, $\Lambda_{QCD} \simeq 300$ MeV, one can integrate this
particle out and define a low-energy effective field theory applicable below
this scale.  Evidently, even though $m_s$ is not $\gg \Lambda_{QCD}$, there is
still a nearly complete decoupling. A similar situation of
near-ideal mixing occurs for the $J^{PC}=2^{++}$ tensor mesons
$f_2(1275)$, $f_2'(1525)$ and the $J^{PC}=3^{--}$  mesons
$\omega_3(1670)$, $\phi_3(1850)$ and can also be understood in terms of approximate decoupling of the light $u \bar u + d \bar d$ state from the heavier $s \bar s$ state.

There exist two different types of nonets for $J^P=1^+$ axial-vector mesons which arise as orbitally excited quark-antiquark bound states: $1 \, {}^3P_1$ and $1 \, {}^1P_1$. These two nonets have
different $C$ quantum numbers for their respective neutral mesons, namely $C=+$ and $C=-$.   The non-strange axial
vector mesons, for example, the neutral $a_1(1260)$ and $b_1(1235)$ cannot mix
because of their opposite $C$-parities. In contrast, the mesons $K_{1A}$ and $K_{1B}$, the strange partners of  $a_1(1260)$ and $b_1(1235)$, respectively, do mix to form corresponding physical mass eigenstates $K_1(1270)$ and $K_1(1400)$.  This complicates the
analysis of the mixings of the SU(3)-singlet and SU(3)-octet mesons in the $1
\, {}^3P_1$ and $1 \, {}^1P_1$ nonets.  Various phenomenological studies indicate that the $K_{1A}$-$K_{1B}$ mixing angle $\theta_{K_1}$ is around  either $35^\circ$ or $55^\circ$, but there is no consensus as to whether this angle is greater or less than $45^\circ$.

In the preset work, we shall show that when applying the approximate decoupling of the light $q\bar q$ state from the heavier $s \bar s$ state to the axial-vector mesons, we are able to pin down the mixing angle $\theta_{K_1}$. This is based on the observation that when the $f_1(1285)$-$f_1(1420)$ mixing angle $\theta_{^3P_1}$ and the $h_1(1170)$-$h_1(1380)$ mixing angle $\theta_{^1P_1}$ are determined from the mass relations, they depend on the masses of $K_{1A}$ and $K_{1B}$, which in turn depend on $\theta_{K_1}$. Nearly complete decoupling will allow us to discriminate different solutions of $\theta_{^3P_1}$ and $\theta_{^1P_1}$ and pick up the right mixing angle $\theta_{K_1}$.

The layout of the present paper is organized as follows. We first recapitulate in Sec. II the main results derived in \cite{CS} for isoscalar meson mixing. Then we proceed to consider the mixing of axial-vector mesons in Sec. III and discuss the physical implications in Sec. IV. We give the conclusions in Sec. V.

\section{Some relations for meson mixing}

In this section we recapitulate some results in \cite{CS} for meson mixing, where we have considered an ${\rm SU}(N_c)$ QCD-like theory with $\ell=N_f-1$ massless or light quarks  $q_i$, $i=1,\cdots,\ell$, and one quark $Q$ of substantial mass $m_Q$.
For $N_c=3$ and $\ell=2$, this  theory
is a rough approximation to real QCD, since the current-quark masses of
the $u$ and $d$ quarks satisfy $m_u, \ m_d \ll \Lambda_{QCD}$, while $m_s
\sim 100$ MeV is smaller than, but comparable to $\Lambda_{QCD}$, and one can
focus on the effective QCD theory with the heavy quarks $c, \ b, \ t$
integrated out. Considering the
mass-squared matrix $M^2$ in the basis of the SU(3) singlet and
octet flavor eigenstates $|V_1
\rangle$ and $|V_8\rangle$, respectively, with $|V_1
\rangle=(u\bar u+d\bar d+s\bar s)/\sqrt{3}$ and $|V_8
\rangle=(u\bar u+d\bar d-2s\bar s)/\sqrt{6}$,
we can write it as the real, symmetric matrix\footnote{Here and below,  we follow the common practice of using the
squared masses of the mesons rather than the masses themselves, because for
bosons it is the squared mass that appear in effective Lagrangians.
}
\beq
M^2 = \left( \begin{array}{cc}
    m_1^2     & \delta \\
    \delta    & m_8^2    \end{array} \right ) \ .
\label{msq}
\eeq
This mass matrix is diagonalized according to
\beq
R(\theta) M^2 R(\theta)^{-1} = M^2_{diag.}
\label{rmrinv}
\eeq
with
\beq
M^2_{diag.} = \left( \begin{array}{cc}
     m_L^2 & 0   \\
       0   & m_H^2  \end{array} \right ) \ , \qquad R(\theta) = \left( \begin{array}{cc}
    \cos\theta & \sin\theta   \\
   -\sin\theta & \cos\theta  \end{array} \right ) \ .
\label{msqdiag}
\eeq
The eigenvalues of $M^2$ are given by
\beq
m_{H,L}^2 = \frac{1}{2}\bigg [ m_8^2+m_1^2 \pm \sqrt{(m_8^2-m_1^2)^2
+ 4\delta^2} \ \bigg ] \ .
\label{mhl}
\eeq
One can work backward from the observed masses and mixing
angle to determine $\delta$.
The mass squared matrix then becomes
\beq \label{massM}
M^2 =\left( \begin{array}{cc}
    m_L^2+m_H^2-m_8^2     & -\Big(m_8^2(m_L^2+m_H^2-m_8^2)-m_L^2 m_H^2\Big)^{1/2} \\
    -\Big(m_8^2(m_L^2+m_H^2-m_8^2)-m_L^2 m_H^2\Big)^{1/2}  & m_8^2   \end{array} \right ) \ ,
\eeq
where the mass squared of the SU(3)-octet $m_8^2$ can be determined from the Gell-Mann Okubo mass relation \cite{GMO}.
The mixing angle can be expressed in several different but equivalent forms:
\beq \label{tan2theta}
\tan2\theta  = -\frac{2\delta}{m_8^2-m_1^2} \ ,  \quad
\cos2\theta  = \frac{m_8^2-m_1^2}{m_H^2-m_L^2} \ , \\
\eeq
\beq \label{tantheta}
\tan \theta  = \frac{m_{8}^2-m_H^2}{\delta} \ ,  \qquad
\cot \theta  = -\frac{m_{8}^2-m_L^2}{\delta} \ , \\
\eeq
\beq \label{tanthetasq}
\tan^2\theta  = \frac{ m_{H}^2-m_8^2}{m_8^2-m_L^2} \ , \quad~
\cos^2\theta  = \frac{ m_{8}^2-m_L^2}{m_H^2-m_L^2} \ .
\eeq
Eqs. (\ref{tan2theta}) and (\ref{tantheta}) have the advantage that the magnitude and the sign of the mixing angle are fixed simultaneously.

Applying the Applequist-Carazzone decoupling theorem \cite{ac}, we infer that when $m_s$ is treated as a variable and increases
past $\Lambda_{QCD}$, it is possible to define an effective low-energy theory
with the $s$ quark integrated out.  Hence, the mixing of the meson flavor
eigenstates must be such as to produce a mass eigenstate composed of the light
$u$ and $d$ quarks, and an orthogonal mass eigenstate composed
only of the $s$ quark.  The decoupling angle is given by
\beq
\theta_{\rm dec.} = \arctan \Big ( \frac{1}{\sqrt{2}} \, \Big ) = 35.26^\circ
\ .
\label{theta_ell2}
\eeq
It turns out that the physical mixing angles $39.0^\circ$, $29.5^\circ$ and $32.0^\circ$, respectively, for $J^{PC}=1^{--}$ (vector), $2^{++}$ (tensor) and $3^{--}$ mesons \cite{pdg} are indeed close to the ideal one. Especially, the vector mixing angle is in good agreement with the value $\theta_{V,ph} = (38.58 \pm 0.09)^\circ$ obtained from a recent global fit by KLOE \cite{kloe}.  {\it A priori}, one does not expect a large decoupling effect because $m_s$ is not large compared to $\Lambda_{QCD}$,
$m_s/\Lambda_{QCD} \simeq 1/3$.  Indeed, one of the most intriguing
aspects of $\omega$-$\phi$ mixing is how close this is to the decoupling limit
even though $m_s/\Lambda_{QCD}$ is not $\gg 1$.

\section{Mixing of axial-vector mesons}

In the quark model, two nonets of $J^P=1^+$ axial-vector mesons are expected as the orbital excitation of the $q\bar q$ system.  In terms of the spectroscopic notation $^{2S+1}L_J$, there are two types of $P$-wave axial-vector mesons, namely, $^3P_1$ and $^1P_1$.  These two nonets have distinctive $C$ quantum numbers for the corresponding neutral mesons, $C=+$ and $C=-$, respectively.  Experimentally, the $J^{PC}=1^{++}$ nonet consists of $a_1(1260)$, $f_1(1285)$, $f_1(1420)$ and $K_{1A}$, while the $1^{+-}$ nonet contains $b_1(1235)$, $h_1(1170)$, $h_1(1380)$ and $K_{1B}$.  The non-strange axial vector mesons, for example, the
neutral $a_1(1260)$ and $b_1(1235)$ cannot have mixing because of
the opposite $C$-parities. On the contrary, $K_{1A}$ and $K_{1B}$ are not the physical mass eigenstates $K_1(1270)$ and $K_1(1400)$ and they are mixed together due to the strange and non-strange light quark mass difference. Following the common convention we write\footnote{
The sign of the mixing angle $\theta_{K_1}$ and the relative signs of the decay constants as well as form factors for $K_{1A}$ and $K_{1B}$ were often very confusing in the literature.  As stressed in Ref.~\cite{CK0708}, the sign of $\theta_{K_1}$ is intimately related to the relative sign of the $K_{1A}$ and $K_{1B}$ states which can be arbitrarily assigned.  This sign ambiguity can be removed by fixing the relative sign of the decay constants of $K_{1A}$ and $K_{1B}$. For example,
in the covariant light-front quark model \cite{CCH} and in pQCD \cite{Lu:BtoP}, the decay constants $f_{K_{1A}}$ and $f_{K_{1B}}$ are of opposite sign, while the $D(B)\to K_{1A}$ and $D(B)\to K_{1B}$ transition form factors have the same signs.
Then $\theta_{K_1}$ is positive as the negative one is ruled out by the data of $D^+\to\bar K_1^0(1270)\pi^+$, $D^0\to K^-_1(1270)\pi^+$ \cite{Cheng:DAP,Cheng:DSATP} and also
by the measurements of $B\to K_1(1270)\gamma$ and
$B\to K_1(1400)\gamma$ \cite{Cheng:Kstargamma}.
In this work, we shall choose the convention for decay constants in such a way that $\theta_{K_1}$ is positive.  Therefore, the values of $\theta_{K_1}$ cited from various references below are always positive in our convention.  Note that for the antiparticle states
$\bar K_1(1270)$, $\bar K_1(1400)$, $\bar K_{1A}$ and $\bar K_{1B}$, the mixing angle is of opposite sign to that defined in Eq. (\ref{K1mixing}).
}
\beqs \label{K1mixing}
 \left( \begin{array}{c}
    |K_1(1270) \rangle \\
    |K_1(1400)   \rangle \end{array} \right ) =
\left( \begin{array}{cc}
     \sin\theta_{K_1} & \cos\theta_{K_1}   \\
   \cos\theta_{K_1}  & -\sin\theta_{K_1} \end{array} \right )
    \left( \begin{array}{c}
                 |K_{1A} \rangle \\
                 |K_{1B} \rangle \end{array} \right ) \ .
\eeqs

There exist several estimations on the mixing angle $\theta_{K_1}$ in the literature.  From the early experimental information on masses and the partial rates of $K_1(1270)$ and $K_1(1400)$, Suzuki found two possible solutions $\theta_{K_1}\approx 33^\circ$ and $57^\circ$ \cite{Suzuki}.  A similar constraint $35^\circ\lsim \theta_{K_1}\lsim 55^\circ$ was obtained in Ref.~\cite{Goldman} based solely on two parameters: the mass difference between the $a_1(1260)$ and $b_1(1235)$ mesons and the ratio of the constituent quark masses.  An analysis of $\tau\to K_1(1270)\nu_\tau$ and $K_1(1400)\nu_\tau$ decays also yielded the mixing angle to be $\approx 37^\circ$ or $58^\circ$ \cite{Cheng:DAP}.
Another determination of $\theta_{K_1}$ comes from the $f_1(1285)$-$f_1(1420)$ mixing angle $\theta_{^3\!P_1}$  to be introduced shortly below which can be reliably estimated from the analysis of the radiative decays $f_1(1285)\to \phi\gamma, \rho^0\gamma$ \cite{Close:1997nm}. A recent updated analysis yields $\theta_{^3\!P_1}=(19.4^{+4.5}_{-4.6})^\circ$ or $(51.1^{+4.5}_{-4.6})^\circ$ \cite{KCYang}.\footnote{From the same radiative decays, it was found $\theta_{^3\!P_1}=(56^{+4}_{-5})^\circ$ in \cite{Close:1997nm}. This has led some authors (e.g. \cite{DMLi}) to claim that  $\theta_{K_1}\sim 59^\circ$. However, another solution, namely, $\theta_{^3\!P_1}=(14.6^{+4}_{-5})^\circ$ corresponding to a smaller $\theta_{K_1}$, was missed in \cite{Close:1997nm}.
}
As we shall see below, the mixing angle $\theta_{^3\!P_1}$ is correlated to $\theta_{K_1}$. The corresponding $\theta_{K_1}$ is found to be $(31.7^{+2.8}_{-2.5})^\circ$ or $(56.3^{+3.9}_{-4.1})^\circ$. Therefore, all the analyses yield a mixing angle $\theta_{K_1}$ in the vicinity of either $35^\circ$ or $55^\circ$.

However, there is no consensus as to whether $\theta_{K_1}$ is greater or less than $45^\circ$.
It was found in the non-relativistic quark model that $m^2_{K_{1A}}<m^2_{K_{1B}}$ \cite{DMLi,Burakovsky,Chliapnikov} and hence $\theta_{K_1}$ is larger than $45^\circ$ (see Eq. (\ref{K1Amass}) below).\footnote{As pointed out in \cite{DMLi}, the solutions $\theta_{K_1}=(37.3\pm3.2)^\circ$ obtained in \cite{Burakovsky} and $(31\pm4)^\circ$ in \cite{Chliapnikov} should be replaced by $\pi/2-\theta_{K_1}$.
}
Interestingly, $\theta_{K_1}$ turned out to be of order $34^\circ$ in the relativized  quark model of \cite{Godfrey}.
Based on the covariant light-front model \cite{CCH}, the value of $51^\circ$ was found by the analysis of \cite{Cheng:2009ms}. From the study of $B\to K_1(1270)\gamma$ and $\tau\to K_1(1270)\nu_\tau$ within the framework of light-cone QCD sum rules, Hatanaka and Yang advocated that $\theta_{K_1}=(34\pm13)^\circ$ \cite{Hatanaka:2008xj}. In short, there is a variety of different values of the mixing angle cited in the literature. It is the purpose of this work to pin down $\theta_{K_1}$.

We next consider the mixing of the isosinglet $^3P_1$ states, $f_1(1285)$ and $f_1(1420)$, and the $1^1P_1$ states, $h_1(1170)$ and $h_1(1380)$:
\begin{eqnarray}
 \left( \begin{array}{c}
    |f_1(1285) \rangle \\
    |f_1(1420)   \rangle \end{array} \right ) =
\left( \begin{array}{cc}
     \cos\theta_{^3\!P_1} & \sin\theta_{^3\!P_1}   \\
   -\sin\theta_{^3\!P_1}  & \cos\theta_{^3\!P_1} \end{array} \right )
    \left( \begin{array}{c}
                 |f_1 \rangle \\
                 |f_8 \rangle \end{array} \right ) \ ,
\end{eqnarray}
and
\begin{eqnarray}
 \left( \begin{array}{c}
    |h_1(1170) \rangle \\
    |h_1(1380)   \rangle \end{array} \right ) =
\left( \begin{array}{cc}
     \cos\theta_{^1\!P_1} & \sin\theta_{^1\!P_1}   \\
   -\sin\theta_{^1\!P_1}  & \cos\theta_{^1\!P_1} \end{array} \right )
    \left( \begin{array}{c}
                 |h_1 \rangle \\
                 |h_8 \rangle \end{array} \right ) \ ,
\end{eqnarray}
where $f_1=(u\bar u+d\bar d+s\bar s)/\sqrt{3}$, $f_8=(u\bar u+d\bar d-2s\bar s)/\sqrt{6}$, and likewise for $h_1$ and $h_8$.
Using the squared mass matrix  Eq. (\ref{massM}) with some appropriate replacements such as $m_L=m_{f_1(1285)}$, $m_H=m_{f_1(1420)}$ etc. for $^3P_1$ states and $m_L=m_{h_1(1170)}$, $m_H=m_{h_1(1380)}$ etc. for $^1P_1$ states, and applying the Gell-Mann Okubo relations for the mass squared of the octet states
\beqs
m_8^2(^3\!P_1) &\equiv& m_{^3\!P_1}^2={1\over 3}(4m_{K_{1A}}^2-m_{a_1}^2), \nonumber \\
m_8^2(^1\!P_1) &\equiv& m_{^1\!P_1}^2={1\over 3}(4m_{K_{1B}}^2-m_{b_1}^2),
\eeqs
we obtain from Eqs. (\ref{tantheta}) and (\ref{tanthetasq}) that
\begin{eqnarray} \label{tantheta:A}
\tan\theta_{^3\!P_1}&=& \frac{m_{^3\!P_1}^2-  m_{f'_1}^2} {\sqrt { m_{^3\!P_1}^2(m_{f_1}^2+m_{f'_1}^2- m_{^3\!P_1}^2)-m_{f_1}^2m^2_{f'_1}}}
                             \,, \nonumber\\
\tan\theta_{^1\!P_1}&=& \frac{ m_{^1\!P_1}^2-  m_{h'_1}^2} {\sqrt { m_{^1\!P_1}^2(m_{h_1}^2+m_{h'_1}^2-m_{^1\!P_1}^2)-m_{h_1}^2m^2_{h'_1}}}
                             \,,
\end{eqnarray}
and
\begin{eqnarray} \label{tanthetasq:Aa}
\tan^2\theta_{^3\!P_1}&=&\frac{4 m_{K_{1A}}^2-m_{a_1}^2-3 m_{f'_1}^2}
                             {-4 m_{K_{1A}}^2+m_{a_1}^2+3 m_{f_1}^2}\,, \nonumber\\
 \tan^2\theta_{^1\!P_1}&=&\frac{4 m_{K_{1B}}^2-m_{b_1}^2-3 m_{h'_1}^2}
                             {-4 m_{K_{1B}}^2+m_{b_1}^2+3 m_{h_1}^2}\,,
\end{eqnarray}
where $f_1$ and $f'_1$ ($h_1$ and $h'_1$) are the short-handed notations for $f_1(1285)$ and $f_1(1420)$ ($h_1(1170)$ and $h_1(1380)$), respectively, and
\begin{eqnarray} \label{K1Amass}
 m_{K_{1A}}^2 &=& m_{K_1(1400)}^2 \cos^2\theta_{K_1} + m_{K_1(1270)}^2
 \sin^2\theta_{K_1} \,, \nonumber \\
  m_{K_{1B}}^2 &=&
 m_{K_1(1400)}^2 \sin^2\theta_{K_1} + m_{K_1(1270)}^2 \cos^2\theta_{K_1} \,.
\end{eqnarray}

It is clear that the mixing angles $\theta_{^3\!P_1}$ and $\theta_{^1\!P_1}$ depend on the masses of $K_{1A}$ and $K_{1B}$ states, which in turn depend on the $K_{1A}$-$K_{1B}$ mixing angle $\theta_{K_1}$. Table \ref{tab:axial} exhibits the values of $\theta_{^3\!P_1}$ and $\theta_{^1\!P_1}$ calculated using Eq. (\ref{tantheta:A}) for some representative values of $\theta_{K_1}$.
We see that while $\theta_{^3\!P_1}$ is not far from the ideal mixing angle for $\theta_{K_1}<50^\circ$, $\theta_{^1\!P_1}$ is very sensitive to $\theta_{K_1}$: Its deviation from exact decoupling increases with the increasing $\theta_{K_1}$.

\begin{table}[t]
\caption{The values of the $f_1(1285)$-$f_1(1420)$ and $h_1(1170)$-$h_1(1380)$ mixing angles $\theta_{^3\!P_1}$ and $\theta_{^1\!P_1}$, respectively, calculated using Eq. (\ref{tantheta:A}) for some representative  $K_{1A}$-$K_{1B}$ mixing angle $\theta_{K_1}$.  } \label{tab:axial}
\begin{center}
\begin{tabular}{|c| c c c c  |}
\hline
~~$\theta_{K_1}$~~ & $57^\circ$ & $51^\circ$ & $45^\circ$ & $34^\circ$ \\
 \hline
 $\theta_{^3\!P_1}$ & $52.0^\circ$ & $45.1^\circ$ & $37.9^\circ$ & $23.1^\circ$ \\
 $\theta_{^1\!P_1}$ & ~~$-17.5^\circ$~~ & ~~$-9.1^\circ$~~ & ~~$14.4^\circ$~~ & ~~$28.0^\circ$~~ \\
 \hline
\end{tabular}
\end{center}
\end{table}

In the literature it is often to use Eq. (\ref{tanthetasq:Aa}) to determine the magnitude of the mixing angles $\theta_{^3\!P_1}$ and $\theta_{^1\!P_1}$ and the following relations
\begin{eqnarray}\label{tantheta:Aa}
 \tan\theta_{^3\!P_1}
&=& \frac{4m_{K_{1A}}^2-m_{a_1}^2-3m_{f'_1}^2}
{2\sqrt{2}(m_{a_1}^2-m_{K_{1A}}^2)}\,, \qquad
 \tan\theta_{^1\!P_1}
=\frac{4m_{K_{1B}}^2-m_{b_1}^2-3m_{h'_1}^2}
{2\sqrt{2}(m_{b_1}^2-m_{K_{1B}}^2)}\,
\end{eqnarray}
to fix their signs (see e.g. \cite{pdg,ChengAP}). Consider the squared
mass matrices
\beqs \label{Msq:QM}
M^2(^3P_1) &=& {1\over 3} \left( \begin{array}{cc}
    2m_{K_{1A}}^2+m_{a1}^2+a_{1A}   & -2\sqrt{2}(m_{K_{1A}}^2-m_{a_1}^2) \\
    -2\sqrt{2}(m_{K_{1A}}^2-m_{a_1}^2)  & 4m_{K_{1A}}^2-m_{a_1}^2    \end{array} \right ) \ , \nonumber \\
M^2(^1P_1)&=& {1\over 3} \left( \begin{array}{cc}
    2m_{K_{1B}}^2+m_{b_1}^2+a_{1B}     & -2\sqrt{2}(m_{K_{1B}}^2-m_{b_1}^2) \\
    -2\sqrt{2}(m_{K_{1B}}^2-m_{b_1}^2)  & 4m_{K_{1B}}^2-m_{b_1}^2    \end{array} \right ) \ ,
\eeqs
for $^3P_1$ and $^1P_1$ states, respectively, where
$a_{1A}$ and $a_{1B}$ are the parameters to be introduced below which will be set to zero for the moment. The above squared mass matrices
can be derived from the non-relativistic quark model.
Naively, if we substitute the above mass matrix elements in Eqs. (\ref{tanthetasq}) for $\tan^2\theta$ and (\ref{tantheta}) for $\tan\theta$ and take $m^2_H=m_{f'_1}^2$ ($m_{h'_1}^2$) and $m^2_L=m_{f_1}^2$ ($m_{h_1}^2$) for the $^3P_1$ ($^1P_1$) states, we will obtain
Eqs. (\ref{tanthetasq:Aa}) and (\ref{tantheta:Aa}) for the mixing angles $\theta_{^3\!P_1}$ and $\theta_{^1\!P_1}$.
However, the mixing angles determined from these two equations are not the same in magnitude. For example, $|\theta_{^3\!P_1}|=23.1^\circ$ is deduced from the former and $\theta_{^3\!P_1}=10.5^\circ$ from the latter for $\theta_{K_1}=34^\circ$.
Since Eqs. (\ref{tanthetasq}) and (\ref{tantheta}) are equivalent, one may wonder why the resultant mixing angles are so different. This can be traced back to the fact that the mass eigenvalues $m_H$ and $m_L$ derived from the mass matrices (\ref{Msq:QM}) are not identical to the physical masses of $f_1(1420)$ and $f_1(1285)$, respectively, for $^3P_1$ states and $h_1(1380)$ and $h_1(1170)$ for $^1P_1$ states. That is, the mass matrices (\ref{Msq:QM}) do not lead to  Eqs. (\ref{tanthetasq:Aa}) and (\ref{tantheta:Aa}). \footnote{It should be stressed that Eq. (\ref{tantheta:Aa}) (see also Eq. (14.9) of the Particle Data Group \cite{pdg}) cannot be derived from
any mass matrix. Unlike Eq. (\ref{tantheta:A}),  it is {\it not} equivalent to Eq. (\ref{tanthetasq:Aa}).}
Instead, they lead to the ideal mixing $\theta_{^3\!P_1}=\theta_{^1\!P_1}=35.26^\circ$ (see \cite{CS} for a detailed discussion). In other words, the mass matrices $M^2(^3P_1)$ and $M^2(^1P_1)$ can be diagonalized by the orthogonal rotation matrix
\beq
R(\theta_{\rm dec.}) = \left( \begin{array}{cc}
    \sqrt{\frac{2}{3}} & \frac{1}{\sqrt{3}}   \\
   -\frac{1}{\sqrt{3}}    &  \sqrt{\frac{2}{3}}
 \end{array} \right ) \ .
\label{rmatrixdec}
\eeq
This result is unphysical, since it predicts
that there is a complete decoupling of the $s$ quark regardless of how small
the nonzero mass different $m_s-m_q$ is.  This unphysical result shows that the initial quark model for the mass matrix is too simplistic. To remedy this defect, one takes account of
the fact that there is a propagator correction (for both the kinetic and mass
squared terms) in which the SU(3) flavor-singlet state
$|V_1\rangle$ annihilates to an intermediate virtual purely gluonic
state and then goes back to itself again \cite{Scadron}.  This annihilation process denoted by $a_{1A}$ and $a_{1B}$ cannot
occur for the flavor-SU(3) octet state, $|V_8\rangle$.

Since the squared mass matrix (\ref{Msq:QM}) derived from the non-relativistic quark model is only an approximation, in this work we should rely on the exact squared mass matrix given in (\ref{massM})  to get the mixing angles, namely, those shown in Table \ref{tab:axial}.
We would like to stress once again that Eqs. (\ref{tantheta:A})  and (\ref{tanthetasq:Aa}) all yield the same magnitude for $\theta_{^3\!P_1}$ and $\theta_{^1\!P_1}$, but the former has the advantage that the sign and magnitude of the mixing angles can be fixed simultaneously.

\section{Discussion}

The values of the $f_1(1285)$-$f_1(1420)$ and $h_1(1170)$-$h_1(1380)$ mixing angles $\theta_{^3\!P_1}$ and $\theta_{^1\!P_1}$, respectively, listed in Table \ref{tab:axial} for some representative  $K_{1A}$-$K_{1B}$ mixing angle $\theta_{K_1}$ are the key results of this work.
Although $\theta_{K_1}$ is unknown, we shall argue that the values of $\theta_{^3\!P_1}\sim 23^\circ$ and $\theta_{^1\!P_1}\sim 28^\circ$ as depicted in Table \ref{tab:axial} are strongly preferred for the following reasons:

\begin{enumerate}

\item
As discussed in Sec. II, nearly ideal mixing occurs for vector, tensor and $3^{--}$ mesons. Except for pseudoscalar mesons where the axial anomaly plays a unique role, this feature should also hold  for axial-vector mesons.
It is obvious from Table \ref{tab:axial}  that
the mixing of isosinglet axial-vector mesons is close to the ideal one for $\theta_{K_1}\sim 34^\circ$ and far away from the decoupling limit (especially, $\theta_{^1\!P_1}\sim -18^\circ$) when $\theta_{K_1}\sim 57^\circ$.

\item
Since only the modes $h_1(1170)\to\rho\pi$ and $h_1(1380)\to K\bar K^*,\bar KK^*$ have been seen so far, this implies that the quark content is primarily $s\bar s$ for $h_1(1380)$ and $q\bar q$ for $h_1(1170)$. Likewise, $K^*\bar K$ and $K\bar K\pi$ are the dominant modes of
$f_1(1420)$ whereas $f_1(1285)$ decays mainly to the $\eta\pi\pi$ and $4\pi$
states. These suggest that the quark content is primarily $s\bar
s$ for $f_1(1420)$ and $q\bar q$ for $f_1(1285)$. Therefore, the observed strong decays of isoscalar axial-vector mesons suggest that their mixings are close to nearly decoupling. This in turn implies that $\theta_{K_1}\sim 34^\circ$ is much more favored. Indeed, if $\theta_{K_1}=57^\circ$, we will have $\theta_{^1P_1}=-18^\circ$ and $h_1(1170)=0.60n\bar n-0.80s\bar s$ and $h_1(1380)=0.80n\bar n+0.60s\bar s$ with $n\bar n=(u\bar u+d\bar d)/\sqrt{2}$. It is obvious that the large $s\bar s$ content of $h_1(1170)$ and $n\bar n$ content of $h_1(1380)$ cannot explain why only the strong decay modes $h_1(1170)\to\rho\pi$ and $h_1(1380)\to K\bar K^*,\bar KK^*$ have been seen thus far.

\item
The $f_1(1285)$-$f_1(1420)$ and $h_1(1170)$-$h_1(1380)$ mixing angles $\alpha_{^3\!P_1}$ and $\alpha_{^1\!P_1}$, respectively, in the flavor basis were recently calculated by the Hadron Spectrum Collaboration based on lattice QCD \cite{Dudek:2011tt}. The results are $\alpha_{^3\!P_1}=\pm(31\pm2)^\circ$ and $\alpha_{^1\!P_1}=\pm(3\pm1)^\circ$. Since $\alpha$ is related to the singlet-octet mixing angle $\theta$ by the relation $\theta=35.3^\circ+\alpha$,\footnote{This is different from the relation  $\theta=\alpha-54.74^\circ$ used in \cite{Dudek:2011tt}. Note that the mixing angle $\theta_{^3P_1}=(56^{+4}_{-5})^\circ$ obtained in \cite{Close:1997nm} cannot be deduced from $\alpha_{^3P_1}=(21\pm5)^\circ$ quoted in \cite{Dudek:2011tt} through the latter relation.}
we have the two-fold solutions: $\theta_{^3\!P_1}=(4.3\pm2)^\circ$ or $(66.3\pm2)^\circ$ and
$\theta_{^1\!P_1}=(32.3\pm1)^\circ$ or $(38.3\pm1)^\circ$. Evidently, the value of $\theta_{^1P_1}\sim -18^\circ$ for $\theta_{K_1}\sim 57^\circ$ disagrees sharply with the lattice result. As for $\theta_{^3P_1}$, we recall that a study of the radiative decays $f_1(1285)\to \phi\gamma, \rho^0\gamma$ yields a direct determination of  $\theta_{^3\!P_1}$ to be $(19.4^{+4.5}_{-4.6})^\circ$ or $(51.1^{+4.5}_{-4.6})^\circ$ \cite{KCYang}. Therefore, there is a discrepancy of around  $15^\circ$ between the lattice and phenomenological results. An improved lattice calculation of $\theta_{^3P_1}$ will be desired.

\end{enumerate}

In short, we conclude that $\theta_{^3\!P_1}\approx 23^\circ$ and $\theta_{^1\!P_1}\approx 28^\circ$ are strongly preferred as they are close to the ideal mixing and much favored by the phenomenological analysis.
This in turn implies the preference of $\theta_{K_1}\sim 34^\circ$ over $57^\circ$.

\section{Conclusions}

Various phenomenological studies indicate that the $K_{1A}$-$K_{1B}$ mixing angle $\theta_{K_1}$ lies in the vicinity of $35^\circ$ or $55^\circ$, but there is no consensus as to whether this angle is greater or less than $45^\circ$.
The values of the $f_1(1285)$-$f_1(1420)$ and $h_1(1170)$-$h_1(1380)$ mixing angles $\theta_{^3\!P_1}$ and $\theta_{^1\!P_1}$, respectively, are summarized in Table \ref{tab:axial} for some representative $\theta_{K_1}$
as they depend on the masses of $K_{1A}$ and $K_{1B}$, which in turn depend on the mixing angle $\theta_{K_1}$.
The approximate decoupling of the light $q\bar q$ state from the heavier $s \bar s$ state, which is empirically successful for vector, tensor and $3^{--}$ mesons, should be also valid for other isoscalar mesons except for the pseudoscalar ones. When applying this nearly complete decoupling to axial-vector mesons, we are able to discriminate different solutions of $\theta_{^3\!P_1}$ and $\theta_{^1\!P_1}$ and pick up $\theta_{K_1}\sim 35^\circ$ over $55^\circ$. For $\theta_{K_1}\sim 55^\circ$, the predicted $\theta_{^1P_1}$ disagrees sharply with the recent lattice calculation and the large $s\bar s$ content of $h_1(1170)$ and $q\bar q$ content of $h_1(1380)$ cannot explain the observation of their strong decays. Therefore,
we conclude that $\theta_{K_1}$ is smaller than $45^\circ$ and that $\theta_{^3P_1}\sim 23^\circ$ and $\theta_{^1P_1}\sim 28^\circ$.

\vspace{1cm}
\section*{Acknowledgments}
We are grateful to the hospitality of C. N. Yang Institute for Theoretical Physics at SUNY Stony Brook and to Robert Shrock and Kwei-Choi Yang for valuable comments.  This research was supported in part by the National Science Council of Taiwan, R.~O.~C.\ under Grant No. NSC-100-2112-M-001-009-MY3.

\end{document}